\documentclass[preprint,12pt]{elsarticle}  
\usepackage{graphicx} 
\usepackage{amssymb} 
\usepackage{amsmath} 

\journal{Physics Letters B} 

\begin{document} 

\begin{frontmatter} 

\title{Chirally motivated $K^-$ nuclear potentials} 

\author[a]{A.~Ciepl\'{y}} 
\author[b]{E.~Friedman} 
\author[b]{A.~Gal\corref{cor1}} 
\cortext[cor1]{Corresponding author: Avraham Gal, avragal@vms.huji.ac.il} 
\author[a]{D.~Gazda} 
\author[a]{J.~Mare\v{s}} 
\address[a]{Nuclear Physics Institute, 25068 \v{R}e\v{z}, Czech Republic} 
\address[b]{Racah Institute of Physics, The Hebrew University, 91904 
Jerusalem, Israel} 
                     
\begin{abstract} 
In-medium subthreshold ${\bar K}N$ scattering amplitudes calculated 
within a chirally motivated meson-baryon coupled-channel model are used 
self consistently to confront $K^-$ atom data across the periodic table. 
Substantially deeper $K^-$ nuclear potentials are obtained compared to 
the shallow potentials derived in some approaches from threshold ${\bar K}N$ 
amplitudes, with ${\rm Re}\:V_{K^-}^{\rm chiral}=-(85\pm 5)$~MeV at nuclear 
matter density. When ${\bar K}NN$ contributions are incorporated 
phenomenologically, a very deep $K^-$ nuclear potential results, 
${\rm Re}\:V_{K^-}^{\rm chiral+phen.}=-(180\pm 5)$~MeV, in agreement with 
density dependent potentials obtained in purely phenomenological fits to 
the data. Self consistent dynamical calculations of $K^-$--nuclear quasibound 
states generated by $V_{K^-}^{\rm chiral}$ are reported and discussed. 
\end{abstract} 


\end{frontmatter}

\section{Introduction} 
\label{intro} 

The interaction of $K^-$ mesons with hadronic systems, ranging from $K^-$ 
atoms through $K^-$--nuclear clusters to dense strange hadronic matter 
realized perhaps in neutron stars, is of topical interest \cite{npa804}. 
SU(3) chiral symmetry combined with coupled channel techniques provides 
a theoretical framework in which low energy meson-baryon observables can be 
systematically evaluated. In such an approach the $\Lambda(1405)$ resonance, 
dominating the $\bar K N -\pi\Sigma$ physics at energies near the $\bar K N$ 
threshold, is generated dynamically. It is natural then to expect a strongly 
attractive and absorptive near-threshold $K^-$--nuclear interaction which 
might support $K^-$--nuclear clusters \cite{weise10} and $K^-$ condensation 
in compact stars \cite{SSS08}. A typical scale for the nuclear-matter depth 
of chirally motivated $K^-$--nuclear potentials is 100~MeV \cite{weise96}, 
although considerably shallower potentials have also been derived \cite{RO00}. 
Outside of chirally motivated interaction models, suggestions were put 
forward for much stronger potentials that could lead eventually to fairly 
narrow quasibound $K^-$--nuclear clusters once the strong transition 
${\bar K} N \to\pi\Sigma$ is kinematically forbidden \cite{AY02}. 
These suggestions have stimulated experimental searches, the best known of 
which claiming rather broad signals of a deeply bound $K^-pp$ configuration, 
at and below the $\pi\Sigma N$ threshold \cite{FINUDA05,DISTO10}. If these 
claims are substantiated in dedicated experiments, it would appear unavoidable 
to conclude that the $K^-$--nuclear potential depth is considerably greater 
than 100~MeV.{\footnote{The distinction between deep ($\gtrsim 150$~MeV) 
and shallow ($\lesssim 50$~MeV) $K^-$--nuclear potentials has been 
recently discussed within $\Lambda$ hypernuclear formation rate 
calculations \cite{CFGK10} which were found to slightly favor 
a deep potential interpretation of recent FINUDA spectra \cite{FINUDA10}.}} 
It would also imply a rather strong two-nucleon nonmesonic absorption mode 
$\bar K NN \to YN$ ($Y\equiv \Lambda,\Sigma$), considerably beyond 
a chiral-model estimate $\Gamma\approx 22$~MeV for a single-nucleon 
induced $\Lambda(1405) N \to YN$ conversion \cite{SJK09}. 
In the present Letter we focus on the use of in-medium chirally motivated 
$K^-$--nuclear potentials in kaonic atoms and discuss the constraints 
provided by a self consistent analysis of kaonic atom data. 

Strong interaction level shifts and widths in kaonic atoms present 
a unique source of information on the $K^-$--nuclear interaction at 
threshold \cite{FG07}. Phenomenological analyses of large data sets 
encompassing the whole periodic table, using optical potentials, reveal 
characteristic features of the interaction. These could reflect on the 
underlying $K^-N$ interaction in the medium, in particular its energy and 
density dependence. Phenomenological density-dependent $K^-N$ scattering 
amplitudes allow for very good fits to kaonic atom data \cite{FGB93,MFG06} 
but the depths of the real optical potential are typically up to 
four times larger than what some in-medium chiral models predict at 
threshold \cite{RO00}. Another feature of empirical kaonic atom optical 
potentials is that the real part is {\it compressed} relative to the 
corresponding nuclear density, with r.m.s. radius smaller than the nuclear 
r.m.s. radius. The reverse is true for the imaginary part. It is shown below 
that these are natural consequences of the density dependence of the present 
chiral model in-medium amplitudes. 

The Klein Gordon (KG) dispersion relation satisfied by $K^-$ mesons in medium 
of density $\rho$ is of the form  
\begin{equation} 
\omega_K^2 - {{\vec p}_K}^{~2} - m_K^2 - \Pi_K(\omega_K,\rho) = 0,    
\label{eq:KG} 
\end{equation} 
where $\Pi_K(\omega_K,\rho)=2({\rm Re}\:\omega_K)V_{K^-}$ is the self energy 
(SE) operator for a $K^-$ meson with momentum ${\vec p}_K$ and energy 
$\omega_K$. A ${\vec p}_K$ momentum dependence 
of $\Pi_K$ is suppressed since, as is shown below, it may be transformed into 
density and energy dependence. In terms of the in-medium $K^-N$ c.m. forward 
scattering amplitude $F_{K^-N}(\sqrt{s},\rho)$ (here assumed $s$-wave): 
\begin{equation} 
\Pi_K(\omega_K,\rho) \approx -4\pi (1+\frac{\omega_K}{m_N})F_{K^-N}
(\sqrt{s},\rho)\rho, 
\label{eq:V} 
\end{equation} 
where $s=(E_K+E_N)^2-({\vec p}_K+{\vec p}_N)^2$ is the Lorentz invariant 
Mandelstam variable $s$ which reduces to the square of the total $K^-N$ 
energy in the two-body c.m. frame and $m_N$ is the nucleon mass. In the 
laboratory frame, $E_K=\omega_K$. The KG dispersion relation (\ref{eq:KG}) 
in bound state applications for a $K^-$ meson leads to a KG equation 
satisfied by the $K^-$ wavefunction \cite{FG07}: 
%\begin{eqnarray} 
%&[\:\nabla^2 - 2\mu({\cal B}_K+V_c) + (V_c+{\cal B}_K)^2 & \nonumber \\ 
%&+ 4\pi(1+\frac{m_K-B_K}{m_N})F_{K^-N}(\sqrt{s},\rho)\:\rho\:]\:\psi=0, & 
%\label{eq:waveq} 
%\end{eqnarray} 
\begin{equation} 
[\:\nabla^2 - 2\mu({\cal B}_K+V_c) + (V_c+{\cal B}_K)^2  
+ 4\pi(1+\frac{m_K-B_K}{m_N})F_{K^-N}(\sqrt{s},\rho)\:\rho\:]\:\psi=0, 
\label{eq:waveq} 
\end{equation} 
where $\mu$ is the $K^-$-nucleus reduced mass, $V_c$ is the Coulomb 
potential generated by the finite-size nuclear charge distribution, 
and ${\cal B}_K= B_K + {\rm i}{\Gamma_K}/2$ is a complex binding energy, 
including a strong interaction width $\Gamma_K$. Finite-medium corrections 
are applied in our calculations specifically to the coefficient of $F_{K^-N}$ 
in Eq.~(\ref{eq:waveq}). 

Wycech \cite{wycech71} and Bardeen and Torigoe \cite{BT72} suggested 
long time ago within $\Lambda(1405)$-based phenomenological models that 
{\it subthreshold} $K^-N$ scattering amplitudes are relevant in kaonic atom 
studies, where the kaon energy is essentially at threshold. In the present 
Letter we construct the $K^-$ meson SE operator (\ref{eq:V}) near threshold 
($\omega_K \approx m_K$) from in-medium subthreshold $K^-N$ scattering 
amplitudes derived within a chirally motivated meson-baryon coupled-channel 
separable-potential model \cite{CS10}. It is shown for the first time how the 
energy and density dependence of $F_{K^-N}(\sqrt{s},\rho)$ leads to a deep 
$K^-$--nuclear potential $V_{K^-}$ in kaonic atoms. This conclusion holds 
already at the lowest-order Weinberg-Tomozawa level that provides excellent 
starting point for modern chiral models \cite{weise10}. We also report 
on self-consistent calculations of $K^-$ quasibound nuclear states based 
on in-medium extensions of the free-space model of Ref.~\cite{CS10}. 
The calculated binding energies and widths are compared with similar entities 
calculated in Ref.~\cite{WH08}.

\section{Subthreshold energy and density dependence} 
\label{sec:V_K} 

The present calculations are based on the chirally motivated meson-baryon 
coupled-channel separable-potential model of Ref.~\cite{CS10}. This free-space 
model expands consistently and systematically to next-to-leading-order (NLO) 
within the heavy baryon formulation of chiral perturbation theory. The 
low-energy constants of the model are fitted to low energy $K^-p$ scattering 
and reaction data. The free-space $K^-N$ scattering amplitude $F_{K^-N}$ is 
given then in a separable form:  
\begin{equation} 
F_{K^-N}=g(k)f_{K^-N}(\sqrt{s})\:g(k^\prime), \,\,\,\,\,\, 
g(k)=\frac{\alpha^2}{k^2+\alpha^2}, 
\label{eq:F} 
\end{equation} 
with c.m. momenta ${\vec k},{\vec k^\prime}$. Of the several versions 
specified in Table 3 of Ref.~\cite{CS10}, we chose the one with a range 
parameter $\alpha=639$~MeV.{\footnote{This version, labeled below CS30, 
produces two $I=0$ $S$-matrix poles in the neighborhood of the $\bar KN$ 
threshold, the lower one at Re~$\sqrt{s}=1398$~MeV evolves from the 
$\pi\Sigma$ channel, the upper one at 1441~MeV -- from the $\bar K N$ channel, 
within the range of values provided by NLO chiral calculations, e.g., 
Ref.~\cite{BMN06}.}} Other chirally motivated potential models that start 
with zero range, i.e. no momentum dependence, in practical applications 
often introduce at least implicitly a finite range, e.g., a cutoff momentum 
$q_{\rm max}=630$~MeV \cite{RO00} or a renormalization scale $\mu=630$~MeV 
\cite{HW08}. The momentum dependence of the amplitude $F_{K^-N}$ in the 
separable-potential model of Ref.~\cite{CS10} is rather weak for the 
applications discussed in the present work and is secondary to the strong 
energy dependence generated by the $\Lambda(1405)$ resonance. 

\begin{figure}[thb] 
\begin{center} 
\includegraphics[height=7cm]{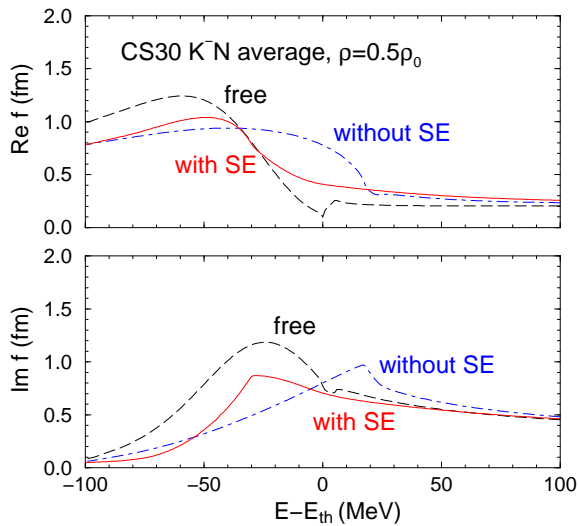} 
\caption{Energy dependence of the c.m. $K^-N$ reduced amplitude 
(\ref{eq:isoscalar}) in version CS30 of the chiral nodel \cite{CS10} below 
and above $E_{\rm th}=m_K+m_N=1432$~MeV. Dashed curves: free-space amplitude; 
dot-dashed curves: Pauli blocked amplitude at $0.5\rho_0$; solid curves: 
including meson and baryon self energies (SE), also at $0.5\rho_0$.} 
\label{fig:CS30} 
\end{center} 
\end{figure}

A typical resonance-shape energy dependence is shown in Fig.~\ref{fig:CS30} 
by dashed lines (marked `free') for the {\it reduced} scattering amplitude 
\begin{equation} 
f_{K^-N}(\sqrt{s}) = \frac{1}{2}[\:f_{K^-p}(\sqrt{s})+f_{K^-n}(\sqrt{s})\:]  
\label{eq:isoscalar} 
\end{equation} 
which is appropriate for the interaction of $K^-$ mesons with symmetric 
nuclear matter. The imaginary part peaks about 25~MeV below the $\bar{K}N$ 
threshold, and the real part rapidly varies there from weak attraction above 
to strong attraction below threshold. While $f_{K^-N}(\sqrt{s})$ at and near 
threshold is constrained by data that serve to determine the parameters 
of the chiral model, the extrapolation to the subthreshold region suffers 
from ambiguities and depends on the applied model \cite{weise10}. 
The free-space reduced scattering amplitude shown in Fig.~\ref{fig:CS30} 
is quantitatively similar to the corresponding $K^-N$ scattering amplitudes 
in other chirally motivated models exhibited in Fig.~7 of Ref.~\cite{HW08}.  

Also shown in Fig.~\ref{fig:CS30} is the energy dependence of two in-medium 
versions of $f_{K^-N}(\sqrt{s},\rho = 0.5\rho_0)$, where $f_{K^-N}(\sqrt{s},
\rho=0) \equiv f_{K^-N}(\sqrt{s})$. One version, in dot-dashed lines (marked 
`without SE'), follows Ref.~\cite{weise96} to require Pauli blocking in the 
intermediate $\bar K N$ states for $\rho \neq 0$. The resulting $f_{K^-N}$ 
exhibits now a resonance-like behavior about 20 MeV above threshold, in 
agreement with Ref.~\cite{weise96}. The other version, in solid lines (marked 
`with SE'), follows Ref.~\cite{CFGM01} to add self consistently meson and 
baryon self energies in intermediate states, similarly to Refs.~\cite{RO00,
lutz98}. The resulting in-medium $f_{K^-N}$ is strongly energy dependent 
below threshold, with a resonance-like behavior about 30~MeV below threshold. 
Similar results are obtained at full nuclear matter density $\rho_0=0.17$ 
fm$^{-3}$. We note that whereas the two in-medium reduced amplitudes shown in 
the figure are close to each other far below and far above threshold, they 
differ substantially at and near threshold. This applies also to the full 
amplitudes since the form factors $g(k)$ remain intact in the transition from 
free-space to in-medium separable amplitudes. At threshold, in particular, 
the real part of the `with SE' amplitude is about half of that `without SE', 
corresponding to a depth $-{\rm Re}~V_{K^-}(\rho_0)\approx 40\!-\!50$~MeV, 
in agreement with Ramos and Oset \cite{RO00}. 

The c.m. reduced amplitudes in Fig.~\ref{fig:CS30} are functions of 
$\sqrt{s}$. In the two-body c.m. system ${\vec p}_K+{\vec p}_N = 0$, but in 
the $K^-$--nucleus c.m. system (approximately nuclear laboratory system) 
${\vec p}_K+{\vec p}_N \neq 0$. Averaging over angles yields 
$({\vec p}_K+{\vec p}_N)^2 \to (p_K^2+p_N^2)$. For bound hadrons we expand 
near threshold, neglecting quadratic terms in the binding energies 
$B_K=m_K-E_K$, $B_N=m_N-E_N$: 
\begin{equation} 
\sqrt{s} \approx E_{\rm th} - B_N - B_K - \xi_N\frac{p_N^2}{2m_N} - 
\xi_K\frac{p_K^2}{2m_K}, 
\label{eq:sqrts} 
%\end{eqnarray} 
\end{equation} 
where $\xi_{N(K)}=m_{N(K)}/(m_N+m_K)$. 
For the square of the relative momenta $\vec k$, $\vec k^\prime$ in form 
factors $g$, Eq.~(\ref{eq:F}), we again average on angles, yielding  
\begin{equation} 
k^2, ~{k^\prime}^2 \rightarrow \xi_N\xi_K(2m_K\frac{p_N^2}{2m_N}+
2m_N\frac{p_K^2}{2m_K}). 
\label{eq:k^2} 
\end{equation} 
Replacing in Eqs.~(\ref{eq:sqrts}) and (\ref{eq:k^2}) the kinetic energy 
$p_K^2/(2m_K)$ in the local density approximation by 
$-B_K -{\rm Re}\:{\cal V}_{K^-}(\rho)$ where ${\cal V}_{K^-}=V_{K^-}+V_c$, 
and approximating the nucleon kinetic energy $p_N^2/(2m_N)$ in the Fermi 
gas model by $23\,(\rho/\rho_0)^{2/3}$~MeV, Eq.~(\ref{eq:sqrts}) becomes 
\begin{equation} 
{\sqrt{s}} \approx E_{\rm th} - B_N - \xi_N B_K - 
15.1(\frac{\rho}{\rho_0})^{2/3}+\xi_K{\rm Re}\:{\cal V}_{K^-}(\rho).  
\label{eq:sfinal} 
\end{equation} 
The downward energy shift invoked by Eq.~(\ref{eq:sfinal}), with respect 
to $E_{\rm th}$, corresponds to that implied in the impulse approximation 
when the many-body $\bar{K}N$ amplitude evaluated at the threshold energy 
$E_{\rm th}$ is replaced by the two-body amplitude at the c.m. energy 
$\sqrt{s}$. Note its density dependence. In most of the subsequent discussion 
it is used as is, although we also checked the effect of implementing gauge 
invariance through the substitution $\sqrt{s} \to \sqrt{s}-V_c$. 
We note that gauge invariance was not implemented in the solution of the 
free-space Lippmann-Schwinger equations for the underlying chiral model 
of Ref.~\cite{CS10}, nor in its in-medium extension displayed in 
Fig.~\ref{fig:CS30}. 

For completeness, we comment on the range of momenta $k,k^\prime$ transcribed 
by Eq.~(\ref{eq:k^2}). In naive applications to kaonic atoms where one 
assumes $p_K \approx 0$, these momenta are due to Fermi motion, as given 
by the first term on the r.h.s. of Eq.~(\ref{eq:k^2}) which is bounded by 
$k(\rho_0),k^\prime(\rho_0) \sim 72$~MeV, quite negligible compared to 
the momentum dependence scale $\alpha=639$~MeV in model CS30. However, 
for strongly attractive $K^-$ nuclear potentials, reaching depths of about 
180~MeV in phenomenological studies \cite{FG07}, the second term on the 
r.h.s. of Eq.~(\ref{eq:k^2}) dominates, yielding $k(\rho_0),k^\prime(\rho_0) 
\sim 276$~MeV. These momenta are nonnegligible, but they are well within the 
CS30 momentum dependence scale $\alpha=639$~MeV which emerged by fitting to 
$K^-p$ low-energy data.

\section{Kaonic atoms} 
\label{sec:atoms} 

We now apply the method outlined above to the interaction of $K^-$ mesons 
with nuclei close to threshold, as a means of testing the subthreshold chiral 
amplitude formalism against experimental results. A distinction was made in 
solving the KG equation (\ref{eq:waveq}) between proton and neutron densities, 
replacing $F_{K^-N}(\sqrt{s},\rho)\rho(r)$ by 
${\cal F}_{K^-N}^{\rm eff}(\sqrt{s},\rho)\rho(r)$, where 
\begin{equation} 
{\cal F}_{K^-N}^{\rm eff}(\sqrt{s},\rho)\rho(r)=
F_{K^-p}(\sqrt{s},\rho)\rho_p(r)+F_{K^-n}(\sqrt{s},\rho)\rho_n(r), 
\label{eq:Feff} 
\end{equation} 
with $\rho_p$ and $\rho_n$ normalized to $Z$ and $N$, respectively, 
and $Z+N=A$. The reduced amplitudes $f_{K^-p}$ and $f_{K^-n}$ were 
evaluated at $\sqrt{s}$ given by Eq.~(\ref{eq:sfinal}), where the atomic 
binding energy $B_K$ was neglected with respect to $B_N\approx 8.5$~MeV. 
A similar approximation was made in Eq.~(\ref{eq:k^2}) for 
$k^2,{k^{\prime}}^2$. $B_K$ was also neglected with respect to $m_K$ in 
Eq.~(\ref{eq:waveq}). The corresponding potential was calculated at each 
radial point for every target nucleus in the data base. Self consistency was 
required in solving Eq.~(\ref{eq:sfinal}) with respect to ${\rm Re}\:V_{K^-}$, 
i.e., the value of ${\rm Re}\:V_{K^-}(\rho)$ in the expression for $\sqrt s$ 
and in the form factors $g$ had to agree with the resulting 
${\rm Re}\:V_{K^-}(\rho)$. 

\begin{figure}[thb]  
\begin{center} 
\includegraphics[height=7cm]{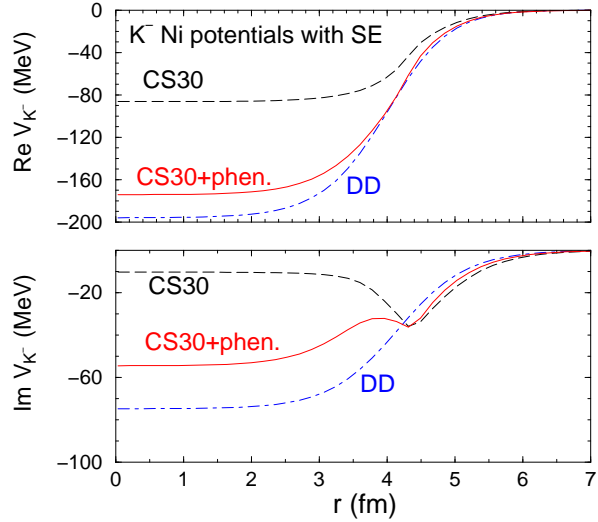} 
\caption{$K^-$--nuclear potentials for $K^-$ atoms of Ni. Dashed curves: 
derived self consistently from in-medium CS30 amplitudes; solid curves: 
plus phenomenological terms from global fits; dot-dashed curves: 
purely phenomenological DD potentials from global fits.} 
\label{fig:atpotl} 
\end{center} 
\end{figure} 

Figure \ref{fig:atpotl} shows, as representative examples, several $K^-$--Ni 
potentials. The CS30 `with SE' amplitudes, within the self consistent 
procedure described above and without adjustable parameters, yield 
the potential marked CS30. For ${\rm Re}\:V_{K^-}$, similar depths 
to within a few MeV are obtained using CS30 `without SE' amplitudes. 
${\rm Re}\:V_{K^-}^{\rm CS30}$ is twice deeper, $-85$~MeV with respect 
to $-40$~MeV, than the shallow potential (not shown here) used in the 
kaonic atom calculations of Ref.~\cite{BGN00}. That shallow potential 
followed from a {\it threshold} value $f_{K^-N}(E_{\rm th},\rho)$, 
without going subthreshold. Yet, ${\rm Re}\:V_{K^-}^{\rm CS30}$ is 
not as deep as ${\rm Re}\:V_{K^-}^{\rm DD}$, where the density dependent 
(DD) potential $V_{K^-}^{\rm DD}$, also shown in the figure, represents 
the best fit with $\chi ^2=103$ for 65 data points obtained using purely 
phenomenological DD potentials \cite{FG07}. 

\begin{figure}[htb] 
\begin{center} 
\includegraphics[height=5cm,width=6cm]{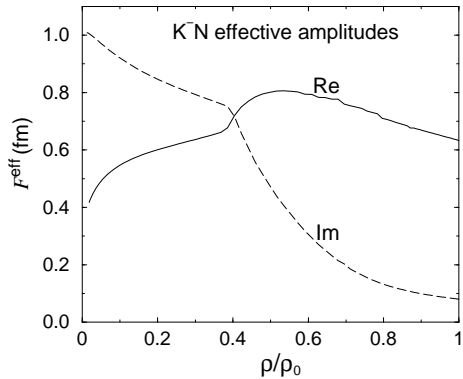} 
\caption{Density dependence of the in-medium `with SE' CS30 self consistent 
subthreshold amplitude ${\cal F}_{K^-N}^{\rm eff}$ for Ni.} 
\label{fig:effamp}
\end{center} 
\end{figure}

The density dependence of the chiral CS30 `with SE' effective scattering 
amplitude ${\cal F}_{K^-N}^{\rm eff}$, calculated self consistently for Ni, 
is shown in Fig.~\ref{fig:effamp}. The increase of ${\rm Re}\:{\cal F}_{K^-N}
^{\rm eff}(\rho)$ with density over the nuclear surface region combined with 
the decrease of ${\rm Im}\:{\cal F}_{K^-N}^{\rm eff}(\rho)$ is the underlying 
mechanism behind the compression and inflation of successful phenomenological 
best-fit potentials such as $V_{K^-}^{\rm DD}$ \cite{FGB93}. However, the 
decrease of ${\rm Im}\:{\cal F}_{K^-N}^{\rm eff}$ is unreasonably rapid, 
leading in the lower part of Fig.~\ref{fig:atpotl} to the peculiar shape of 
${\rm Im}\:V_{K^-}^{\rm CS30}$ at the nuclear surface.{\footnote{No similar 
bulge at the nuclear surface appears if ${\rm Im}\:V_{K^-}^{\rm CS30}$ 
is derived from CS30 `without SE' amplitudes, and 
${\rm Im}\:V_{K^-}^{\rm CS30}(\rho_0)$ is then about twice that for the 
`with SE' case shown in the figure.}} A substantial $K^-N$ energy shift into 
the subthreshold region is involved in the calculation of 
${\cal F}_{K^-N}^{\rm eff}$, e.g., $-40$~MeV at $0.5\rho_0$ down to $-60$~MeV 
at $\rho_0$ in Ni. For such far subthreshold energies, the reduced amplitude 
${\rm Im}\:f_{K^-N}^{\rm CS30}$, as seen in Fig.~\ref{fig:CS30}, is too small 
to provide the absorptivity required by the kaonic atom data. Note that these 
single-nucleon amplitudes do not account for multi-nucleon absorption which 
becomes increasingly important at subthreshold energies. 

Figure~\ref{fig:atpotl} also demonstrates the effect of adding adjustable 
$\rho$ and $\rho^2$ terms to $V_{K^-}^{\rm CS30}$, resulting in a best-fit 
potential $V_{K^-}^{\rm CS30+phen.}$ with $\chi ^2$ of 164. For the imaginary 
part, the depth $-{\rm Im}\:V_{K^-}^{\rm CS30}(\rho_0)$ is substantially 
increased towards reaching $-{\rm Im}\:V_{K^-}^{\rm DD}(\rho_0)$, with 
a weaker bulge now at the nuclear surface. For the real part, the depth 
$-{\rm Re}\:V_{K^-}^{\rm CS30}(\rho_0)$ is also significantly increased 
from 85~MeV to 175~MeV, close to the phenomenological potential depth 
$-{\rm Re}\:V_{K^-}^{\rm DD}$. Similar results, to within a few MeV, 
hold when starting from the CS30 `without SE' amplitudes and also when 
the substitution $\sqrt{s}\to\sqrt{s} - V_c$ is made in the r.h.s. of 
Eq.~(\ref{eq:sfinal}). We note that the phenomenological addition to both 
${\rm Im}\:V_{K^-}^{\rm CS30}$ and ${\rm Re}\:V_{K^-}^{\rm CS30}$ is dominated 
by $\rho^2$ terms which are required by the fit procedure and which are 
likely to represent ${\bar K}NN$ absorptive and dispersive contributions, 
respectively. The emerging phenomenology is similar to that for $V_{\pi^-}$ 
in pionic atom studies where theoretically motivated single-nucleon 
contributions are augmented by phenomenological $\rho^2$ terms representing 
$\pi NN$ processes \cite{EE66}. More work is required to justify 
microscopically the size of the $\rho^2$ kaonic atom contributions suggested 
by the successful $V_{K^-}^{\rm CS30+phen.}$. 

Although our study of kaonic atoms focused on $K^-N$ $s$-wave interaction 
models incorporating the subthreshold $\Lambda(1405)$ resonance, we also 
checked whether $p$-wave contributions from the subthreshold $\Sigma(1385)$ 
resonance are sizable and, more importantly, whether they could modify the 
pattern of $V_{K^-}^{\rm CS30}$ and $V_{K^-}^{\rm CS30+phen.}$ established 
above. To this end we used the $\Sigma(1385)$-based $p$-wave potential of 
Weise and H\"{a}rtle \cite{WH08} consisting of resonance and background parts. 
%While detailed account will be given elsewhere, we summarize our preliminary 
%calculations as follows: (i) starting with $V_{K^-}^{\rm CS30}$ and varying 
%only the overall strength of the $p$-wave components in a $\chi^2$ fit 
%resulted in wiping out the resonance part while augmenting the small 
%background part; (ii) varying the overall strength of the $p$-wave component, 
%along with adding phenomenological $s$-wave terms in the construction of 
%$V_{K^-}^{\rm CS30+phen.}$, also resulted in wiping out the $p$-wave 
%resonance part while retaining a small $p$-wave background term. The $s$-wave 
%part of $V_{K^-}^{\rm CS30+phen.}$ remained very close to where it was in the 
%absence of $p$-wave terms. 
While detailed account will be given elsewhere, we assert that fits to 
kaonic atom data within the present self-consistent procedure do not 
require substantial $p$-wave contributions, and that $V_{K^-}^{\rm CS30}$ 
and $V_{K^-}^{\rm CS30+phen.}$ are robust with respect to adding $p$ waves.

\section{$K^-$--nuclear quasibound states} 
\label{sec:nuclear} 

The threshold $K^-$--nuclear potential $V_{K^-}^{\rm CS30}$ shown in 
Fig.~\ref{fig:atpotl} is sufficiently attractive to generate $K^-$--nuclear 
quasibound states. However, since the potential is energy dependent, it has 
to be calculated again self consistently for binding energies $B_K$ in the 
expected range of tens of MeV. Calculations for $^{16}$O and $^{208}$Pb 
were reported by Weise and H\"{a}rtle \cite{WH08} using a subthreshold 
extrapolation 
\begin{equation} 
\sqrt{s}\approx E_{\rm th}-B_K-V_c 
\label{eq:WH} 
\end{equation} 
for solving self consistently the $K^-$ KG equation within a local potential 
approximation of a chiral-model $\bar K N$ amplitude. The present self 
consistency scheme is based on Eq.~(\ref{eq:sfinal}) for $\sqrt{s}$ and 
Eq.~(\ref{eq:k^2}) for $k^2$, as applied here to kaonic atoms, but without 
neglecting $B_K$ for $K^-$ nuclear states. We thus solved the KG equation 
(\ref{eq:waveq}), requiring self consistency explicitly for $B_K$ 
(and implicitly for ${\rm Re}\:V_{K^-}$) using Eqs.~(\ref{eq:k^2}) and 
(\ref{eq:sfinal}). 
Realistic RMF density distributions $\rho(r)$ were employed, within a fully 
dynamical calculation that allows the nuclear density to get polarized by the 
$K^-$ meson. In this dynamical scheme, following Refs.~\cite{MFG06,GFGM07}, 
a RMF self consistency cycle is applied in which the $K^-$ meson solution 
serves further as a source in the RMF equations of motion which are solved to 
produce a modified nuclear density that goes again into the $K^-$ meson KG 
equation. However, the present formulation differs fundamentally from previous 
RMF calculations: here $V_{K^-}$ is generated microscopically from a two-body 
coupled channel chiral model and, furthermore, energy and density dependencies 
are introduced directly through the underlying $K^-N$ scattering amplitude. 

\begin{table}[hbt] 
\begin{center} 
\caption{Binding energies $B_K$ and widths $\Gamma_K$ (in MeV) of 
$1s$ $K^-$ nuclear quasibound states, calculated self consistently 
from Eq.~(\ref{eq:waveq}) using in-medium `with SE' CS30 amplitudes 
$F_{K^-N}(\sqrt{s},\rho)$ within a dynamical RMF scheme \cite{MFG06,GFGM07}. 
Possible $K^-NN \to YN$ $\rho^2$ contributions are excluded.} 
\begin{tabular}{rrrrrr} 
\hline  
 & $^{12}$C & $^{16}$O & $^{40}$Ca & $^{90}$Zr & $^{208}$Pb \\ 
\hline 
%$B_K$ & 53.3 & 53.4 & 67.1 & 76.4 & 81.6 \\ 
$B_K$ & 54.8 & 54.9 & 68.2 & 77.3 & 82.2 \\
$\Gamma_K$ & 11.7 & 11.4 & 9.8 & 9.4 & 8.6 \\ 
%$\Gamma_K$ & 41.4 & 45.5 & 50.1 & 48.5 & 44.4 \\ 
\hline 
\end{tabular} 
\label{tab:B} 
\end{center} 
\end{table} 

Binding energies and widths calculated dynamically for the $1s$ $K^-$ nuclear 
quasibound state in several nuclei across the periodic table are listed in 
Table~\ref{tab:B}. The `with SE' in-medium version of the CS30 chiral model 
was used. Similar results are obtained for the `without SE' version, with 
slightly larger binding energies and widths. The values of $B_K$ listed in 
the table are in close agreement with binding energies calculated within 
a dynamical RMF approach for nucleons and antikaons \cite{GFGM07} when the 
$K^-$ nuclear interaction is mediated exclusively by an $\omega$ vector field, 
with the same pure-$F$ SU(3) coupling used in chiral models. The listed 
$B_K$ values are smaller by 10--30~MeV than those calculated statically in 
Ref.~\cite{WH08}. We checked that applying the gauge-invariant substitution 
$\sqrt{s}\to\sqrt{s}-V_c$ in the r.h.s. of Eq.~(\ref{eq:sfinal}) makes 
little difference to the systematics of the binding energies and widths 
in Table~\ref{tab:B}, increasing $B_K$ gradually between 1~MeV for $^{12}$C 
to 5~MeV for $^{208}$Pb. The real potential depths associated with the binding 
energies obtained here are of order 100~MeV (e.g. 110~MeV for the converged 
${\rm Re}\:V_{K^-}$ in $^{40}$Ca). They follow naturally from the strong 
energy dependence of $f_{K^-N}(\sqrt{s},\rho)$ at and below threshold which is 
incorporated within a genuinely self consistent and dynamical calculation of 
kaonic nuclei. The values of $f_{K^-N}(E_{\rm th},\rho)$, which for in-medium 
`with SE' imply ${\rm Re}\:V_{K^-}(\rho_0)\approx -50$~MeV, are of no 
relevance to the actual binding energies of $K^-$ nuclear quasibound states. 

The calculated widths $\Gamma_K$ listed in Table~\ref{tab:B} represent only 
$K^-N\to \pi Y$ absorption processes that are accounted for by the coupled 
channels chiral model without allowing for multinucleon absorption modes. 
They are relatively small, of order 10~MeV, reflecting the proximity of the 
$\pi\Sigma$ thresholds which suppresses the major decay modes available for 
a $K^-$ meson on a single nucleon. The sensitivity of the width calculation 
to the precise form of in-medium subthreshold extrapolation of $\sqrt{s}$ is 
demonstrated by five separate static calculations of the $K^-$ $1s$ quasibound 
state in $^{16}$O listed in Table~\ref{tab:W}. The first three calculations 
use the CS30 `without SE' version of the in-medium chiral model. The first 
one uses Eq.~(\ref{eq:WH}) for self consistency in the solution of the KG 
equation, mocking up as much as possible within our model the calculation by 
Weise and H\"{a}rtle \cite{WH08}. 
The next two calculations use Eq.~(\ref{eq:sfinal}) which shifts $\sqrt{s}$ 
further down with respect to where Eq.~(\ref{eq:WH}) shifts it, reducing 
$\Gamma_K$ from about 50~MeV to about 20~MeV. The effect of subtracting $V_c$ 
in the r.h.s. of Eq.~(\ref{eq:sfinal}), which yields the ($B_K,\Gamma_K$) 
values of the column marked `Eq.~(\ref{eq:sfinal})$-V_c$', is seen to be 
minor by comparing with the third ($B_K,\Gamma_K$) column. The last two 
calculations also use Eq.~(\ref{eq:sfinal}) for self consistency, but the 
CS30 in-medium version `without SE' is replaced by `with SE' which brings 
the calculated value of $\Gamma_K$ further down to near its value in the 
dynamical calculation reported in Table~\ref{tab:B}. The last column includes 
also a $\Sigma(1385)$ $p$-wave contribution given in Ref.~\cite{WH08}. 
Its effect is found to be rather small since $\sqrt{s}$ is well below the 
peak of the $\Sigma(1385)$ resonance. A more detailed account of $K^-$ 
quasibound state calculations incorporating $p$ waves will be given elsewhere. 

\begin{table}[thb] 
\begin{center} 
\caption{Binding energies $B_K$ and widths $\Gamma_K$ (in MeV) of the $1s$ 
$K^-$ nuclear quasibound state in $^{16}$O, calculated self consistently from 
Eq.~(\ref{eq:waveq}) using in-medium CS30 amplitudes $F_{K^-N}(\sqrt{s},\rho)$ 
within a static RMF scheme for various prescriptions of treating $\sqrt{s}$. 
Unless stated to the contrary, the CS30 version is `without SE'. 
Possible $K^-NN \to YN$ $\rho^2$ contributions are excluded.} 
\begin{tabular}{cccccc} 
\hline 
 & Eq.~(\ref{eq:WH}) & Eq.~(\ref{eq:sfinal})$-V_c$ & 
Eq.~(\ref{eq:sfinal}) & Eq.~(\ref{eq:sfinal})+SE & $+p$ waves \\ 
\hline 
$B_K$ & 58.2 & 53.0 & 51.9 & 51.2 & 54.2 \\ 
$\Gamma_K$ & 49.8 & 21.6 & 19.0 & 11.8 & 12.1 \\ 
\hline 
\end{tabular} 
\label{tab:W} 
\end{center} 
\end{table} 

In addition to the $K^-N \to \pi Y$ single-nucleon induced widths of $K^-$ 
quasibound states which are calculable from given in-medium chiral models and 
which were shown above to be quite model dependent, there are also sizable 
two-nucleon absorption processes $K^-NN \to YN$ with considerably lower 
thresholds that contribute additional widths of order 40~MeV \cite{MFG06,WH08,
GFGM07}. These absorption processes could be simulated by adding energy 
dependent imaginary $\rho^2$ terms suggested by the (CS30+phen.) potential 
for kaonic atoms together with the associated dispersive real $\rho^2$ terms. 
This requires further consideration.

\section{Conclusion} 
\label{sec:concl} 

In conclusion, we have shown within in-medium extensions of the chirally 
motivated coupled-channel separable-interaction CS30 model \cite{CS10} 
how to incorporate the strong energy and density dependence of the $K^-N$ 
scattering amplitude $F_{K^-N}(\sqrt{s},\rho)$, at and below threshold, 
into a self consistent evaluation of the SE operator $\Pi_K(\omega_K,\rho)$ 
for kaonic atoms. The procedure adopted here is sufficiently general to be 
applied within in-medium extensions of other chirally motivated interaction 
models. Our calculations demonstrate that kaonic atom data probe the 
subthreshold regime of the in-medium $K^-N$ scattering amplitude. The 
two in-medium extensions of CS30 considered in the present work produce 
parameter-free $K^-$ nuclear potentials for kaonic atoms which are similar to 
each other, with depths $-{\rm Re}\:V_{K^-}^{\rm CS30}(\rho_0)=(85\pm 5)$~MeV 
and $-{\rm Im}\:V_{K^-}^{\rm CS30}(\rho_0)=(20\pm 10)$~MeV evaluated at the 
center of a typical medium-weight nucleus such as Ni. Preliminary calculations 
using the Tomozawa-Weinberg lowest order chiral interaction term, including 
the new SIDDHARTA $1s$ level shift and width in kaonic hydrogen \cite{SID11} 
produce similar results. The sharp decrease of ${\rm Im}\:f_{K^-N}$ 
towards the $\pi\Sigma$ threshold is reflected in rapid decrease 
of ${\rm Im}\:{\cal F}_{K^-N}^{\rm eff}$ with increased density. 
Therefore absorption processes beyond a single nucleon mechanism 
must be added together with a possible dispersive real term. Indeed to 
achieve truly low $\chi^2$ values for $K^-$ atoms, phenomenological 
potential terms had to be added, leading to increased $V_{K^-}$ depths, 
i.e., $-{\rm Re}\:V_{K^-}^{\rm CS30+phen.}(\rho_0)=(180\pm5)$~MeV 
and $-{\rm Im}\:V_{K^-}^{\rm CS30+phen.}(\rho_0)=(70\pm 20)$~MeV. 
These dominantly $\rho^2$ terms could represent $K^-NN$ processes 
outside of the present single-nucleon chiral model. 

$K^-$ nuclear quasibound states generated by in-medium extensions of CS30 
were also calculated within a self consistent scheme. Potential depths 
$-{\rm Re}\:V_{K^-}(\rho_0)$ of order 100~MeV were obtained for both 
extensions used, exceeding somewhat the depths derived for kaonic atoms. This 
indicates a moderate state, or energy dependence of $\Pi_K(\omega_K,\rho_0)$. 
The similarity of the results for the two in-medium extensions used here is 
in striking contrast to the large difference in their Re~$f_{K^-N}$ values 
at threshold, as can be judged from Fig.~\ref{fig:CS30}. Implementing 
additional phenomenological term within $K^-$ nuclear quasibound calculations 
is likely to result in $B_K$ values higher than 100~MeV. This topic deserves 
further consideration. 

Our calculations provide for the first time a microscopic link between 
shallow $K^-$ nuclear potentials \cite{RO00,BGN00} obtained from threshold 
$K^-N$ interactions and phenomenological deep ones deduced from kaonic atom 
data \cite{FGB93,MFG06}. In future work, $p$-wave $\bar K N$ subthreshold 
amplitudes associated with the $\Sigma(1385)$ resonance will be explored 
in detail, to confirm the secondary role played by these amplitudes in our 
preliminary extended studies of kaonic atoms{\footnote{See also 
Ref.~\cite{GNOR02} for a related calculation using threshold amplitudes.}} 
and in studies of quasibound $K^-$ nuclear states \cite{WH08,GFGM07}.

\section*{Acknowledgements} 
Stimulating discussions with Wolfram Weise are gratefully acknowledged. 
This work was supported by the GACR Grant No. 202/09/1441, as well as 
by the EU initiative FP7, HadronPhysics2, under Project No. 227431.

\end{document}